# Quantum Computers: Engines for Next Industrial Revolution

## English Abstract

Although the current information revolution is still unfolding, the next industrial revolution is already rearing its head. A second quantum revolution based on quantum technology will power this new industrial revolution with quantum computers as its engines. The development of quantum computing will turn quantum theory into quantum technology, hence release the power of quantum phenomena, and exponentially accelerate the progress of science and technology. Building a large-scale quantum computing is at the juncture of science and engineering. Even if large-scale quantum computers become reality, they cannot make the conventional computers obsolete soon. Building a large-scale quantum computer is a daunting complex engineering problem to integrate ultra-low temperature with room temperature and micro-world with macro-world. We have built hundreds of physical qubits already but are still working on logical and topological qubits. Since physical qubits cannot tolerate errors, they cannot be used to perform long precise calculations to solve practically useful problems yet.

# 标题： 量子计算机：下次工业革命的引擎
## 量子计算的现状与前景专题文章约稿

## 摘要

尽管信息革命方兴未艾，下一次工业革命已经萌芽。这次革命将是以量子技术为基础，以量子计算机为引擎的第二次量子革命。其发展将把量子理论转化为量子技术，从而释放出量子现象的潜能，指数级般地提高计算速度和科学技术进步。建造大规模量子计算机现在正处于科学和工程的交接当口。即使大规模量子计算机建成，在可期的未来也不会全面取代现在的计算机。建造大规模量子计算机是一个从微观到宏观，从低温到常温，从科学到技术的复杂工程问题。目前造成的是大约一百个物质量子比特的没有纠错能力的量子计算机。因为不能纠错，所以不能精确持久地解决有实际应用价值的问题。

关健词：

量子计算，量子比特，物质量子比特，逻辑量子比特，拓扑量子比特

## 引言

在人类发展的历史长河中，数次工业革命极大地改善了人类的生活条件。第一次工业革命是以蒸汽机为引擎；第二次主要是以电力为能源的大规模生产模式；现在我们正处在信息革命中，晶体管是我们的引擎。尽管这次工业革命可以说是方兴未艾，但是下一次工业革命已经萌芽。这即将来临的工业革命将是以量子技术为基础，所以实际上量子计算机只是这次量子技术革命当中一个比较突出的技术而已。第一次量子革命是理论的建立，第二次量子革命将把量子理论转化为量子技术，从而释放出量子现象的潜能，指数级般地改善计算速度和科学技术进步。量子计算机首先是一种量子眼镜，正如望远镜和显微镜赋予人类探索遥远星际和微观世界的新能力，量子计算机会让我们感受到量子世界的丰富多彩。建造量子计算机不可避免但又困难重重，从现在到所谓的可扩展通用量子计算机还有很远的路要走。建造可扩展通用量子计算现在应该正处于科学和工程的交接当口。

量子计算是量子信息科学的一部分。我们在本文不会涉及另外部分包括量子通信。广义的计算已经渗入到了社会的每一个层面，从而计算速度和信息储存成为计算机发展的重要颈瓶。量子计算机不仅会大大提高计算速度，也会带来计算理论的一次范式革命。但即使大规模量子计算机建成，在很长时间内也不会全面取代现在的计算机，就像飞机的出现并没有让汽车消失。第一，长期内，量子计算机会非常昂贵；第二，大多数计算现在的计算机已经绰绰有余；第三，量子计算机现在仅仅对某些特殊的问题有超快的算法。

经过科学家几十年的努力，量子计算机的建造终于到了一个拐点时刻：量子霸权的发表(后面解释怎么回事)【1】。但这并不意味着可扩展通用量子计算的出现就在眼前。建造大规模量子计算机是一个从微观到宏观，从低温到常温，从科学到技术的复杂工程问题。量子计算机本身是一个矛盾体，一方面要把量子比特从环境中完全孤立出来，另外要让他

们有可控的相互作用。还有很多科学技术工程方面的困难问题有待解决，但这并没难倒科学家。

量子计算的信息单位是量子比特。量子比特是所有拥有两个经典态的量子系统的数学抽象，所以是具有二维希尔伯特空间量子系统的统称。典型的例子是电子自旋---电子不单可以自旋向上或向下，也可以是它们的任何线性叠加。

量子计算机是很多量子比特的多体系统。描述量子计算机最粗糙的说法是有多少量子比特，但更重要的是量子比特的精确度。另外量子比特又分物质量子比特和逻辑量子比特。物质量子比特没有纠错机制，而逻辑量子比特能够主动在计算过程中不断地纠正足够小的随机错误。物质量子比特的建造就是控制一个物理系统两个经典态的叠加。它的实现五花八门，最领先的是超导和离子阱。逻辑量子比特的实现极其困难。只有在物质量子比特造的足够多和精确时才可以考虑，我们目前已经能够考虑造逻辑量子比特。微软独树一帜：拓扑量子计算不做物质量子比特，因为拓扑量子比特应该等同于逻辑量子比特。但拓扑量子比特与逻辑量子比特思路不同：逻辑量子比特是按量子纠错思路，相当于软件纠错，拓扑量子比特不主动纠错，而是利用拓扑物质形态的特有刚性避免局部错误，计算不受破坏。

目前造成的是大约一百个物质量子比特的没有纠错机制的量子计算机。因为不能纠错，所以不能精确的大规模计算。目前还没有找到解决实际问题的应用。另外还有几百甚至上千个量子比特的量子模拟机，但业内人士并没共识这些模拟机是否利用了量子现象，也不清楚是否超越现有计算机。

1. 量子计算的理论与历史

从四千年前的算盘到现在我们口袋里的手机，信息处理的原理都是基于所谓的比特，也就是像投硬币那样有两个结果的随机过程。量子比特是经典比特在量子力学里的体现，用到线性叠加和纠缠原理，所以理论是完全不同的，这个我们后面讨论。

为什么我们要考虑量子计算？第一，摩尔定律已近机限。经典计算机现在差不多到了经典技术的极限，进入量子世界。现在晶体管的尺寸大约 3 到 7 纳米，在这个尺寸里面量子现象已经非常重要了。芯片技术一个重要问题就是要散热，即使不做量子计算，在如此微观的量子世界里，量子力学必须考虑在内。第二，量子计算显然从理论上看有巨大的超算能力。和现在最好的经典计算办法比，例如分解数因子，量子计算有指数级增长的优势【3】。很多秘密都基于数因子的分解对经典计算机而言是很难的假设。如果存在大规模的量子计算机，这些秘密将不再安全。第三，量子计算机在理论上是可行的。如果量子理论在大规模量子计算理论中的应用还成立，那么理论上量子计算是可以实现的。所以有众多理由让我们想做量子计算。

什么是量子计算？量子计算的想法可以溯源到物理学家费曼（Feynman）1959 年的一个饭后讲话：微观世界还有足够多的余地【2】。随后有很多非常重要的进展，但是真正引起人们注意的是 1994 年数学家秀尔（P. Shor）给出的一个量子分解数因子的高效算法【3】。秀尔的算法证明用量子计算机分解数因子是多项式增长类的复杂性，而不是现在相信的指数增长类的。这是一个非常了不起的计算机算法结果。但随后马上有人指出这个结果只是数学家的定理而已，因为谁也造不出来量子计算机。量子力学有个未知不能复制的原理：一个未知量子态是不可能随便完全复制一份的。这不像我们经常做的，如果拿到一篇即使不懂的数学文章，还是可以去复印一份，这在量子世界里做不到。所以经典计算理论里的简单重复纠错的办法在量子世界里是不可行的。数学家是很聪明的，1995 年秀尔做了另外一项工作，至少跟他的算法一样重要，他发现在量子计算里还是可以有纠错的办法【4】。所以在大约 1996 年的时候基本上已经证明理论上是可以造出大规模量子计算机。量子纠错这个工作非常重要，奠定了造可扩展通用量子计算机的理论基础，我也是从 1996 年开始研究量子计算的。

从某种意义上讲，任何一个量子物理实验室都有一种量子比特，但能否用它做量子计算机则是一个完全不同的问题。谷歌用超导体做了一个 6 乘 9 方块的量子芯片。本来是 54 个量子比特，但有一个坏掉了，所以就是 53 个可以计算的量子比特。他们的文章宣布做到了所谓的量子霸权【1】---证明存在一个量子计算机只需三分种就能解决的问题，而世界上最好最快的超级计算机估计也要一万年。当然科学实验一定要能别人重复，到现在为止别人还没有能重复。假设有别人能重复，那就证明量子计算是一种新的计算模型，而这种新模型是基于量子力学理论的。

毕达哥拉斯说过什么事情都可以用数来代表。这个观点在计算机理论里表现为任何一个问题的解决都可以变成一个函数的计算：输入的是一个数，算出的答案也是一个数。什么是个计算模型呢？每一个物理理论都能给出一个计算模型。选定一个物理理论，然后给我一个数，把这个数变成这种理论里物理系统的一个态，这个物理系统就会随着时间的流失而演化，这个演化的过程就是计算的过程，停下来的时候物理状态给出一个数，就是计算结果。当然停下来的时候要刚好停在给出答案的状态上。所以计算模型就像一个黑匣子，把经典力学放进去就是经典计算模型---图灵机。我们现在用的计算机模型都是基于经典物理。如果在黑匣子放进量子力学，计算模型就是量子计算。

量子现象怪诞不可思议，所以解释各种说法都有，由此而建立的量子力学是一次物理革命。一个重要原因在于量子力学在微观世界里给我们一种新的存在状态的描述。量子力学最重要的原理是量子叠加。在经典世界中，例如要描述我的位置，可以说11月16号下午6点钟王正汉在北京。如果我是活在量子世界中，那我可能在宇宙当中任何一个地方出现。我的量子态就在宇宙里每一个我能出现的地方写下一个复数，全部复数的绝对值平方加起来要等于1。这就是量子力学很不同的地方，即任何经典上可能的状态都可以成为线性叠加中的一员。一个具体例子就是著名的双夹缝实验，如果你躲在有双夹缝屏幕背后，有人用一枝电子枪打你，什么地方是最安全? 如果电子是个经典的粒子的话，显然两个空中间的背后。电子不是经典的，电子是量子的，你认为安全的地方是最不安全的。实验发现这两个缝中间背后是最被容易打中的，这就是因为量子线性叠加原理。双夹缝实验各种解释

都有，但是量子力学给出了最科学的解释。在双夹缝实验里每个电子有两个态，一个态是电子可以从左边的缝过去，另一个态也可以从右边过去，两种可能可以做线性叠加，所以才会出现匪夷所思的现象。谷歌的实验证明可以有2的53次方这么多东西叠加在一起。2的53次方是9007199254740992。这个数字有多大？全世界所有美元加起来都比这个数小的多。在巨大的数字之下量子力学还成立是件了不起的事情。

量子力学还有一个更加怪诞的所谓量子纠缠。爱因斯坦称之为幽灵般的行为。如果一个量子系统有很多子系统，我们是不好说这个子系统处在怎样一个状态之下，因为只有整体才有一个所谓的纯态。 只描述子系统的状态要失掉很多关于整体态的信息，这就是所谓量子纠缠。

量子力学的动态行为也是完全不同的。第一，一个量子态是很多经典态的叠加，所以当时间演变时，所有经典态都同时跟着演化；第二，量子态有所谓的量子隧穿效应。假设有一个粒子，没有足够的能量爬到山顶上， 那么在经典世界里，这粒子就到不了山的另一边。但在量子世界的粒子就有可能过得去。这实际上是时间能量测不准原理一个应用。另外还有所谓隐性传输。一个量子系统的状态一方面是由那些粒子构成，另一方面是这些粒子怎么形成的这个态。量子隐性传输所谓 DNA，也就是这些粒子怎么造出来这个态的的信息，而不是传输这些粒子物质过去。

量子力学的数学描述用得是希尔伯特空间，也就是复线性代数。任何一个物理系统所有可能的态形成一个希尔伯特空间。每一个态对应于希尔伯特空间里面的一条线。时间演化是通过酉算子在态上的作用实现，相当于是薛定鄂方程的解。

用量子计算解决一个问题，就是计算一个函数。首先就是要把输入数表达在一个量子系统状态上，然后按照量子力学时间演化，通过测量结束的态得到计算结果。量子力学测量的时候会出现概率，这就是为什么量子计算结果会是概率性的。量子力学不是我们最精确的物理理论，量子场论更精确。那么把量子场论放在黑匣子里面会得到什么样的计算模型？大多人相信得不到新的计算模型，也就说用量子场论得到的计算模型跟量子力学是一回事，也是量子计算。 这只是一个猜想，是我做量子计算的原因。如果考虑特殊的拓扑量子场论，这是我跟弗里德曼（M. Freedman）和基塔耶夫（A. Kitaev）证明的一个定理【5】。从这这个角度看，也可以理解为什么量子计算困难。现实世界的材料遵循的是量子场论，所以建造量子计算机是在用量子场论模拟量子力学。

为什么量子计算能算的更快？计算复杂性根本上是一个熵的问题，难是因为有很多不同的选择，有很多选择就有所谓的熵。用经典计算机，要找到答案，要把每个可能性都试过去。 在量子力学里，可以并行，会快是有道理的。前面提到了谷歌的有 53 个量子比特的量子计算机实现量子霸权做什么事情？53 个量子比特的量子态是一个巨大数 9007199254740992 这么多经典状态的叠加，每一个状态都有一个复数加权，取这个数绝对值平方就得到一个 9007199254740992 样东西的一个概率分布。谷歌就是采样这样的概率分布，仅用 200 秒，谷歌最好的计算机估计也要 1 万年。IBM 认为大概三天就能做得到。但第一，IBM 这是一个理论结果；第二，要用到巨大的足球场大小的外面储存；第

三，要用世界上最快的经典计算机。即使 IBM 是对的，比这样的计算机快 1000 倍也是某种量子霸权了。谷歌的量子霸权目前还没有实际应用价值，仅仅是科学技术上的进步。

量子计算机有没有什么真正好的应用？前面说过，我个人觉得即使没有别的，仅仅可以让我们对量子世界有真正的感觉这一条就非常值得了。当然还有很多别的可能，从开发新材料，新医药到解决气候问题。

2. 量子计算机的建造与现状

建造量子计算机最大的困难在于量子系统非常脆弱，发生退相干：由于误差的不断积累，量子性丢失，成为经典态。怎样控制误差也成为量子计算成功与否的关健。目前有两种思路：一是计算过程中随时纠错，另外依靠特殊材料能够不受某些误差的坡坏。随之
量子计算建造也分成两派，一派微软独此一家做拓扑量子计算，用到拓扑物质形态；剩下都是分两步走，第一步要造非常精确的所谓物质量子比特，第二步做量子纠错，造出逻辑量子比特。目前物质量子比特最多最好的是谷歌用超导体做的 53 个物质量子比特。目前还没有拓扑量子比特，也没有逻辑量子比特。量子计算要真正改变世界一定要可扩展通用。可扩展是指我们可以在可期的未来不断地增加量子比特的个数。通用是指我们可以通过不同软件解决各种问题。领域共识是可扩展通用量子计算的建造一定需要逻辑量子比特或者拓扑量子比特。

为了区别于拓扑量子计算，我们把分两步造量子计算称为传统量子计算。

传统量子计算的第一步已有将近三十年的历史。进步远远超出很多专家的预期。目前超导量子计算已很接近可以走第二步的精度。超导量子计算准备实现的量子纠错码是一种拓扑纠错码【6】。

除了超导量子计算，还有离子阱，电子自旋，光子等等。超导量子计算也有不同的设计。除了高精度的要求，传统量子计算可扩展也非常艰难。一个逻辑量子比特目前估计要上百个高质量的物质量子比特。

拓扑量子计算的想法来自于我的博士导师弗里德曼和基塔耶夫【7】。拓扑量子计算需要找到非常特殊的拓扑物质形态。如果有，就可以从"真空"里边生成一些非交换任意子，做一个辫子的操作让它们演变。我和弗里德曼和拉森（M. Larsen）证明至少理论上存在一种任意子，可以作通用量子计算【8】。跟别的量子计算比较就在于拓扑量子计算直接进入第二步。为什么拓扑？拓扑实际上是一种整体几何。一个典型的例子是欧拉示性数。每一个曲面都有一个欧拉示性数，是一个整数。这个整数可以写成一个曲率的积分，那个曲率可以随便变，但积分总和必须是整数。这个欧拉示性数描述的就是拓扑性质：局部的可以任何变化，但是整体变不了。因为很多错误、误差都是局部的，但这个拓扑性质是整体的，所以表述在拓扑性质里的信息是免疫的：不会被局部的错误、误差破坏。微软要实现的非交换任意子现象叫马耶拉纳（Majorana）零模。

拓扑量子比特的建造一种是用人造拓扑材料，一种用天然拓扑材料。天然拓扑材料主要是分数量子霍尔效应的二维电子系统【9】。人造的主要是半导体和超导体混合成的纳米线【10】。但目前都没能确定非交换任意子的存在，因此也就还没有拓扑量子比特。如果有，理论上拓扑量子计算是可扩展的。

目前世界主要国家都投入巨大的财力人力建造量子计算机。美国起步很早，在离子阱，超导和拓扑量子计算等领域明显领先。加拿大和澳大利亚几乎和美国同时长期投入，效果显著。欧洲等国家起步晚些，近几年也加大投资。建造通用大规模量子计算机是一个长期的困难的点滴积累的过程，需要全世界默默无闻科学家们坚持不懈地不断辛苦耕耘，而大量人力财力支持也会加快它的步伐。

量子计算从开始就有很多反对的声音，但大多的理由并不科学，属于我怎么都不信一类。另一类是基于计算复杂性理论，但计算复杂性理论并不适用于现实实验，因为现实里的计算机并不完全符合理论模型的机限。我们知道量子态的波函数不可能无限精准---测不准原理，所以可扩展通用量子计算不能实现也是有可能的。如果量子计算不能实现是因为量子理论需要修正，这个结局同样重要。

3. 量子比特里的世界

现实物理世界是量子的，量子态的个数是指数增长的。要深刻认识和控制量子现象需要模拟量子世界，但经典计算机是不可能做到的，因为三百个量子比特用到的经典态的数目已经超出了可见全宇宙里原子的个数。拓扑物质态里经常有上千个电子，它们的量子态就是上千个量子比特。因次只有用量子计算机模拟量子世界才是可行的。

量子技术的潜力无法预测，但毫无疑问会深刻地影响量子计量，量子化学，先进材料，纳米成像等领域。会加速室温超导体的发现，帮助设计无毒染料，提供发明碳汇和改进化肥催化剂的方向。

量子计算随至带来的软件革命同样深刻。全新的软件人才的培养和教育，全新的编程语言，操作系统和框架。

现在的小量子计算机可以当作量子仪器用于量子教育。量子世界与人类属不同世界，只有通过不断地模拟才会产生真正的量子直觉。

基于经典物理的科学技术为人来生活带来了天翻地复的变化。改变人类命运的航空，通信，医药等航业如何继续发展都遭遇到空前的挑战。许多问题的解决都需要全新的办法，量子技术虽然尚未起步，但无疑提供了无数的可能。

量子计算机将成为人类进入量子世界的门户，但是看到的量子世界很难分清是明月光还是地上霜。工具的建造与使用是人类的特性，量子计算机的建造是人类发展中的新篇章。无论量子计算机建造成功与否，量子技术的革命都会极大地推进科学与技术的进步，为人类带来福祉。量子思想的影响也会延神到逻辑等，从而改变数学等自然科学的基础，为人类发掘量子世界的秘密提供新的语言和工具。

文献引用：